\documentclass[onecolumn,showpacs]{revtex4}
\usepackage{amssymb}
\usepackage{makeidx}
\usepackage{amsmath}
\usepackage{graphicx}
\usepackage{epstopdf}
\usepackage{dcolumn}
\usepackage{bm}

\input{tcilatex}
\begin{document}

\title{Microscopic nonlinear quantum theory of absorption of coherent
electromagnetic radiation in doped bilayer graphene}
\author{A.G. Ghazaryan}
\email{amarkos@ysu.am}
\author{Kh.V. Sedrakian}
\affiliation{Centre of Strong Fields, Yerevan State University, 1 A. Manukian, Yerevan
0025, Armenia }
\date{\today }

\begin{abstract}
The microscopic quantum theory of nonlinear stimulated scattering of chiral
particles in doped $AB$ stacked bilayer graphene on Coulomb field of charged
impurities in the presence of strong coherent electromagnetic radiation is
presented. The Liouville-von Neumann equation for the density matrix is
solved analytically. Here the interaction of electrons with the scattering
potential is taken into account as a perturbation. The absorption rate of
nonlinear inverse-bremsstrahlung for a grand canonical ensemble of fermionic
chiral particles is calculated using the obtained solution. The analysis of the
 obtained rate shows that in the terahertz and near-infrared range of
frequencies there is significant absorption of incident radiation via
multiphoton stimulated bremsstrahlung mechanism.
\end{abstract}

\maketitle



\section{Introduction}

Because of the unusual properties of single-layer graphene (SG) \cite%
{1,2,3,4,5,6} and bilayer graphene (BG) \cite%
{1a,2a,KoshinoAndo,2e,3a,3b,3c,Screening,Doped} the parameter of their
interaction with the coherent electromagnetic (EM) radiation is very high in
comparison with other systems. The last can allow the fundamental and
technological applications of multilayer graphene in nano-opto-electronics,
as well as in quantum electrodynamics, low-energy physics, condensed matter
physics and at else \cite%
{7,nanop1,nanop2,7aa,7aaa,Mer2,Mer3,Mer4,Mer5,Mer6,Mer7,Merarxiv1,Merarxiv2,Xu2016}%
. The high absorption coefficient ($\sim $ $10^{6} \mathrm{cm}^{-1}$) \cite%
{trans} already cleared that graphene strongly interacts with light. Hence,
it is very important that because of the gapless structure of intrinsic SG
and $AB$\ stacked BG such interaction can be efficiently realized with the
terahertz or near-infrared devices as high-power generator and frequency
multipliers, as well as a protective material for nanodevices \cite{18}, 
\cite{19}. The absorption of pump wave in nanoscale volumes is a highly
desirable property for shielding materials used in nanoelectronics, the
aerospace industry, where strict requirements exist such as lightness and
smallness or tightness. So, such frequencies are of present interest.

What concerns the absorption of pump wave with the multiphoton stimulated
bremsstrahlung (SB), among the wave-induced processes it is an important
phenomenon of energy exchange between the charged particles and plane
monochromatic wave in plasma-like media \cite{12a}. There are many papers
relating the electrons elastic scattering on impurity ions in SG, with
consideration mainly within the framework of perturbation theory by
electrostatic potential \cite%
{BornEllastic,Chen2008,BornEl1,BornEl2,BornEl3,PhysRev2015,TanAdam2007,Katsnelson,Kaikai Xu17}%
. Regarding the SB process in graphene at moderate intensities of stimulated
radiation, in case of its linear absorption by the charged chiral particles,
at the present time, there are extensive investigations carried out in the
scope of the linear theory, see, e.g. \cite{7b,7b1,7b2,7b3,Zhu2014}.
Multiphoton cross-sections for SB of conduction electrons in intrinsic SG
have been obtained in the Born approximation over the scattering potential
in the presence of an external EM radiation field in \cite{nanop1}, \cite%
{nanop2}.

The interlayer coupling between the two graphene sheets in AB stacked
bilayer graphene strongly change the monolayer's Dirac cone, inducing
trigonal warping on the band dispersion and changing the topology of the
Fermi surface in the low-energy region. Last changes the transport
properties of the BG \cite{1a}, which studied theoretically within a
self-consistent Born approximation in \cite{KoshinoAndo}. The rates of
elastic and transport scattering on charged impurities in SG and BG are
investigated experimentally and comparison with theoretical predictions have been 
made in \cite{Adam,Katsnel,ElastBilayer,Zhang,Sharma,Xiao}. The multiphoton rates and
total cross-sections of the SB process of conductive electrons on the
charged impurity ions in BG had been studied in Born approximation by
scattering potential field in \cite{SBbilayer}.

Meanwhile, the interaction of an electron with the EM wave in plasma media
at the photon energy $\hbar \omega >T_{e}$ is described by dimensionless
intensity parameter of intensity $\chi _{0}=eE_{0}/\omega \sqrt{m_{e}\hbar
\omega }$ \cite{7} ($E_{0}$ is the wave field strength amplitude, $\omega $
is the frequency of the wave, $e$ and $m_{e}$ are the elementary charge and
mass, $T_{e}$ -plasma temperature). The $\chi _{0}$ is the ratio of the
amplitude of the momentum given by the wave field to momentum at the
one-photon absorption. The intensity of the wave expressed by the parameter $%
\chi _{0}$ can be estimated as $I_{\chi _{0}}=\chi _{0}^{2}\times 1.74\times
10^{12}$ $\mathrm{Wcm}^{-2}(\hbar \omega /\mathrm{eV})^{3}$\cite{7aaa}.
Multiphoton effects become essential at $\chi _{0}\sim 1$, which for
terahertz photons $\hbar \omega \sim 0.01$\ $\mathrm{eV}$ corresponds to
intensity $I_{\chi _{0}}\simeq 10^{6}$\ $\mathrm{Wcm}^{-2}$. Meanwhile, in
BG for intraband transitions the wave-particle interaction at the photon
energies $\hbar \omega >\mathcal{E}_{L}$\ \cite{Mer4} characterizes by known
dimensionless intensity parameter $\chi =\chi _{0}(m_{e}\rightarrow m_{\ast
})$ ($m_{\ast }$ is an effective mass of chiral particle, $\mathcal{E}%
_{L}\simeq 1$ $\mathrm{meV}$ is the Lifshitz energy). Hence, for the
realization of multiphoton SB in BG, one can expect $\sim 30$\ times smaller
intensities than for SB in atoms \cite{7aa}, \cite{7aaa}, \cite{12a,12aa,13}.

In the present paper, we have studied the nonlinear absorption coefficient
of external EM radiation in doped BG both analytically and numerically. As a
mechanism for the real absorption or emission of a plane-monochromatic wave
by the charged particles (or plasma-like medium), we have assumed SB process
of conductive chiral particles scattering on the charged impurities in doped
BG. We developed a microscopic quantum theory of graphene nonlinear
interaction with the coherent strong EM radiation of arbitrary. With the
help of the solution of Liouville-von Neumann equation for the density
matrix, we calculated the nonlinear stimulated scattering of chiral
particles in BG on the Coulomb field of impurity ions at the presence of an
external pump radiation, taking into account the interaction of charged
carriers with the scattering potential in the Born approximation.

In Sec. II the quantum dynamics of SB of conductive chiral particles in BG
is presented with analytical results for density matrix and
inverse-bremsstrahlung absorption rates. In Sec. III the analytic formulas
in case of screened Coulomb field of charged impurities are considered
numerically. Conclusions are given in Sec. IV.

\section{The absorption coefficient of coherent EM radiation in doped BG}

We will present the quantum theory of multiphoton SB of charged carriers on
an arbitrary electrostatic potential of impurity ion in doped bilayer
graphene at the presence of the coherent electromagnetic radiation on the
base of the density matrix. The pump wave field is considered exactly, while
the electrostatic scattering potential of doped ions as a perturbation. To
exclude the interband transitions we have considered high Fermi energies and
external EM wave is taken to be in the terahertz or near-infrared domain of
frequencies. In the presented paper, the influence of multiphoton effects in
SB absorption process with an external EM radiation field of moderate
intensities is considered. Note, that the first nonrelativistic treatment of
SB in the Born approximation has been carried out analytically in the work 
\cite{12a}, and then this approach has been extended to the relativistic
domain \cite{12aa}.

The pump wave was applied in the perpendicular direction to the BG sheet ($%
XY $) to omit the effect of the magnetic field. Similar calculations for a
wave linearly polarized along the $OX$ axis show qualitatively the same
picture. The constant phase connected with the position of the wave pulse
maximum with respect to the BG plane is set zero. We assume the radiation
field to be quazimonochromatic and of the linear polarization with amplitude 
$E_{0}$ and frequency $\omega =2\pi /T$:%
\begin{equation}
\mathbf{E}(t)=\widehat{\mathbf{x}}E_{0}\cos \omega t.  \label{0_a}
\end{equation}%
The wave vector potential $\mathbf{A}(t)$ will have the form:%
\begin{equation}
\mathbf{A}(t)=-c\int_{0}^{t}\mathbf{E}(t^{\prime })dt^{\prime }=-\widehat{%
\mathbf{x}}\frac{E_{0}}{\omega }\sin \omega t.  \label{1a}
\end{equation}%
The charged impurities are assumed to be at rest and either randomly or
nonrandomly distributed in the doped graphene, the arbitrary form
electrostatic potential field of which is described by the scalar potential:%
\begin{equation}
\varphi (\mathbf{r})=\sum\limits_{i}^{N_{i}}\phi _{i}(\mathbf{r-R}_{i}),
\label{2a}
\end{equation}%
where $\phi _{i}$ is the potential of a single ion placed at the position $%
\mathbf{R}_{i}$, and $N_{i}$\ is the number of impurity ions in the
interaction region.

We will consider the quantum kinetic equations for a single chiral particle
density matrix for SB process investigation, which can be derived from the
second quantized formalism. As a basis for single-particle wave functions we
take the approximate solution of the charged chiral particle equation in the
strong EM wave field $\mathbf{A}(t)$ we use an even simpler model for BG $AB$
stacked which neglects both the electron-hole asymmetry and the trigonal
warping \cite{2a}, \cite{Screening}, \cite{Nilsson1,Nilsson2}. In accordance
with the nonlinear quantum theory of BG in the mentioned case with energy
window $0.002$ $\mathrm{eV}\lesssim \varepsilon \lesssim 0.1$ $\mathrm{eV}$
in the vicinity of Dirac points $K_{\zeta }$ (valley quantum number $\zeta
=\pm 1$) in the Brillouin zone the fermion particle wave function $\Psi _{%
\mathbf{p}}(\mathbf{r},t)$ in the strong EM wave field may be presented in
the form: 
\begin{equation}
\Psi _{\mathbf{p}}(\mathbf{r},t)=F_{\mathbf{p}}(t)e^{\frac{i}{\hbar }\mathbf{%
pr}}.  \label{2}
\end{equation}%
Hear $\mathbf{r=}\left\{ x,y\right\} $ is the 2D-radius vector, parameter $S$
is the quantization area--graphene layer surface area, 
\begin{equation}
F_{\mathbf{p}}(t)=\frac{1}{\sqrt{2S}}\left( 
\begin{array}{c}
e^{i2\zeta \vartheta } \\ 
\sigma%
\end{array}%
\right) e^{-i\Omega (\mathbf{p},t)},  \label{3}
\end{equation}%
\begin{equation}
\Omega (\mathbf{p},t)=\frac{1}{2m_{\ast }\hbar }\int \left[ \left( p_{x}+%
\frac{e}{c}A_{x}\right) ^{2}+p_{y}^{2}\right] dt,  \label{4a}
\end{equation}%
are the time-dependent spinor wave function $F_{\mathbf{p}\sigma }$ and the
phase $\Omega (\mathbf{p},t)$ (classical action in the field (\ref{1a})), $%
\vartheta (\mathbf{p}+\frac{e}{c}\mathbf{A}(t))$ is the polar angle in
momentum space. BG has a low-energy dispersion which is approximated \cite%
{3b}, \cite{Nilsson2} by massive valence and conduction bands without a gap
(in opposite to massless bands in SG). The quasiparticle energy $\mathcal{E}%
(p)$ is defined by $\mathcal{E}(p)=\sigma \left( p_{x}^{2}+p_{y}^{2}\right)
/\left( 2m_{\ast }\right) $ , where $\sigma =\pm 1$ correspond to the
conduction/valence bands; $m_{\ast }=\gamma _{1}/\left( 2\mathrm{v}%
_{F}^{2}\right) $, $\gamma _{1}\simeq 0.39$ $\mathrm{eV}$ is the interlayer
tunneling amplitude, $\mathrm{v}_{F}$ is the intrinsic SG Fermi velocity.
For the actual parameters, the effective mass for BG is $m_{\ast }\approx
\left( 0.033\div 0.05\right) m_{e}$. The spin and the valley quantum numbers
are conserved. There is no degeneracy upon the valley quantum number $\zeta $%
, for the issue considered. However, since there are no intervalley
transitions, the valley index can be considered as a parameter. The particle
states (\ref{2}) are normalized by the condition%
\begin{equation}
\int \Psi _{\mathbf{p}^{\prime }}^{+}(\mathbf{r},t)\Psi _{\mathbf{p}}(%
\mathbf{r},t)d\mathbf{r}=\frac{(2\pi \hbar )^{2}}{S}\delta \left( \mathbf{p-p%
}^{\prime }\right) .  \label{5a}
\end{equation}

The second quantized Hamiltonian of the system can be presented in the form:%
\begin{equation}
\mathcal{H}=\mathcal{H}_{0}\mathbf{+}\mathcal{H}_{sb}(t).  \label{6a}
\end{equation}%
The first term in Eq. (\ref{6a})\ is the Hamiltonian of a single dressed
chiral particle 
\begin{equation}
\mathcal{H}_{0}=\int \widehat{\Psi }^{+}\widehat{H}_{0}\widehat{\Psi }d%
\mathbf{r},  \label{8a}
\end{equation}%
while the second term $\mathcal{H}_{sb}(t)$ is the interaction Hamiltonian
describing the SB process in the EM field (\ref{1a}) which can have the form:%
\begin{equation}
\mathcal{H}_{sb}(t)=\frac{1}{c}\int \widehat{j}\varphi (\mathbf{r})d\mathbf{%
r,}  \label{8aa}
\end{equation}%
with the current density operator%
\begin{equation}
\widehat{j}=-eg_{s}g_{\mathrm{v}}\int \widehat{\Psi }^{+}(\mathbf{r},t)%
\widehat{\mathbf{v}}\widehat{\Psi }(\mathbf{r},t)d\mathbf{r},  \label{50}
\end{equation}%
where $\widehat{\mathbf{v}}=\partial \widehat{H}/\partial \widehat{\mathbf{p}%
}$ is the velocity operator, $\mathbf{\hat{p}}=\left\{ \widehat{p}_{x},%
\widehat{p}_{y}\right\} $\textbf{\ }is the electron momentum operator, $%
g_{s} $ and $g_{\mathrm{v}}$ are the spin and valley degeneracy factors,
respectively. Here $\widehat{\Psi }$ and $\widehat{H}_{0}$ are the field
operator and the Hamiltonian of the dressed single chiral particle in doped
BG. In particular, for the effective $2\times 2$ Hamiltonian \cite{2a},\cite%
{KoshinoAndo}:

\begin{equation}
\widehat{H}_{0}=\frac{1}{2m_{\ast }}\left( 
\begin{array}{cc}
0 & \left( \zeta \widehat{p}_{x}-i\widehat{p}_{y}\right) ^{2} \\ 
\left( \zeta \widehat{p}_{x}+i\widehat{p}_{y}\right) ^{2} & 0%
\end{array}%
\right) ,  \label{50a}
\end{equation}%
the velocity operator in components reads:

\begin{equation}
\widehat{\mathrm{v}}_{x}=\frac{\zeta }{m_{\ast }}\left( 
\begin{array}{cc}
0 & \left( \zeta \widehat{p}_{x}-i\widehat{p}_{y}\right) \\ 
\left( \zeta \widehat{p}_{x}+i\widehat{p}_{y}\right) & 0%
\end{array}%
\right) ,  \label{51}
\end{equation}%
\begin{equation}
\widehat{\mathrm{v}}_{y}=\frac{i}{m_{\ast }}\left( 
\begin{array}{cc}
0 & -\left( \zeta \widehat{p}_{x}-i\widehat{p}_{y}\right) \\ 
\left( \zeta \widehat{p}_{x}+i\widehat{p}_{y}\right) & 0%
\end{array}%
\right) .  \label{52}
\end{equation}%
Note, that the simplified forms for $\widehat{j}$ (\ref{50}) in doped $AB$
stacked BG in more general cases with taken into account the trigonal
warping and induced gap was obtained in \cite{Mer4}, \cite{Merarxiv1}, \cite%
{Merarxiv2}. For the following, we will make the Fourier transform of scalar
potential in Eq. (\ref{8aa}) 
\begin{eqnarray}
\varphi (\mathbf{r}) &=&\frac{1}{(2\pi )^{2}}\int V(\mathbf{q})e^{-i\mathbf{%
qr}}d\mathbf{q,}  \label{6} \\
V\left( \mathbf{q}\right) &=&\int \sum\limits_{i}^{N_{i}}e\varphi _{i}(%
\mathbf{r-R}_{i})e^{i\mathbf{qr}}d\mathbf{r,}  \notag
\end{eqnarray}%
and the interaction Hamiltonian can be expressed in the form: 
\begin{equation}
\mathcal{H}_{sb}(t)=-\frac{g_{s}g_{\mathrm{v}}}{c(2\pi )^{2}}\int \int 
\widehat{\Psi }^{+}V(\mathbf{q})e^{-i\mathbf{qr}}\widehat{\Psi }d\mathbf{q}d%
\mathbf{r,}  \label{9a}
\end{equation}%
where $\hbar \mathbf{q}=\mathbf{p}^{\prime }-\mathbf{p}$ is the recoil
momentum.

Let's pass to Furry representation and present the Heisenberg field operator
of the chiral particle in the form of an expansion in the quasistationary
states (\ref{2}): 
\begin{equation}
\widehat{\Psi }(\mathbf{r},t)=\int d\Phi _{\mathbf{p}}\widehat{a}_{\mathbf{p}%
}(t)\Psi _{\mathbf{p}}(\mathbf{r},t),  \label{8}
\end{equation}%
where $\widehat{a}_{\mathbf{p}}(t)$ and $\widehat{a}_{\mathbf{p}}^{+}(t)$
are the annihilation and creation operators, respectively, for a chiral
particle with momentum $\mathbf{p}$, associated with positive energy ($%
\sigma =1$) solutions satisfy the anticommutation rules at equal times%
\begin{equation}
\{\widehat{a}_{\mathbf{p}}^{\dagger }(t),\widehat{a}_{\mathbf{p}^{\prime
}}(t^{\prime })\}_{t=t^{\prime }}=\frac{(2\pi \hbar )^{2}}{S}\delta \left( 
\mathbf{p}-\mathbf{p}^{\prime }\right) ,  \label{rul1}
\end{equation}%
\begin{equation}
\{\widehat{a}_{\mathbf{p}}^{\dagger }(t),\widehat{a}_{\mathbf{p}^{\prime
}}^{\dagger }(t^{\prime })\}_{t=t^{\prime }}=\{\widehat{a}_{\mathbf{p}}(t),%
\widehat{a}_{\mathbf{p}^{\prime }}(t^{\prime })\}_{t=t^{\prime }}=0.
\label{rul2}
\end{equation}%
where $d\Phi _{\mathbf{p}}=Sd^{2}\mathbf{p}d\theta \mathbf{/}(2\pi \hbar
)^{2}$ \cite{Ando1982}. With the help of Eqs. (\ref{50})--(\ref{52}) and (%
\ref{8}) the expectation value of the current for the certain valley $\zeta $
in the case of $AB$ stacked BG can be written as 
\begin{equation}
\widehat{j}=-\frac{eg_{s}g_{v}}{\left( 2\pi \hbar \right) ^{2}m_{\ast }}\int 
\mathbf{p}d\mathbf{p}\langle \widehat{a}_{\mathbf{p}}^{+}\left( t\right) 
\widehat{a}_{\mathbf{p}}\left( t\right) \rangle .  \label{53}
\end{equation}%
We have excluded the hole operators in Eqs. (\ref{8})--(\ref{53}), since the
contribution of electron-holes intermediate states will be negligible for
considered EM\ radiation intensities and Fermi energies. Taking into account
anticommutation rules (\ref{rul1}), (\ref{rul2}) and Eqs. (\ref{8}), (\ref{2}%
)--(\ref{4a}), the second quantized Hamiltonian can be expressed in the
form: 
\begin{equation}
H_{0}=\int d\Phi _{\mathbf{p}}\mathcal{E}\left( \mathbf{p}\right) \widehat{a}%
_{\mathbf{p}}^{+}\widehat{a}_{\mathbf{p}}+\mathcal{H}_{sb}(t),  \label{9}
\end{equation}%
where the first term is the Hamiltonian of the wave dressed two-dimensional
chiral particle field, and the second term, which can be expressed by the
form:%
\begin{equation}
\mathcal{H}_{sb}(t)=\int d\Phi _{\mathbf{p}}\int d\Phi _{\mathbf{p}^{\prime
}}T_{\mathbf{p}^{\prime }\mathbf{p}}(t)\widehat{a}_{\mathbf{p}^{\prime }}^{+}%
\widehat{a}_{\mathbf{p}}  \label{11a}
\end{equation}%
is the Hamiltonian of interaction describing the SB of the chiral particle
with the "quasienergy":%
\begin{equation}
\mathcal{E}\left( \mathbf{p}\right) =\frac{\omega }{4\pi m_{\ast }}\int_{0}^{%
\frac{2\pi }{\omega }}\left[ \left( p_{x}+\frac{e}{c}A_{x}(t)\right)
^{2}+p_{y}^{2}\right] dt,  \label{17a}
\end{equation}%
Using the relations (\ref{8}), (\ref{2})-(\ref{4a}) for the impurity
potential of the arbitrary form electrostatic potential $V(\mathbf{q})$ we
can have the following SB amplitudes:

\begin{equation*}
T_{\mathbf{p}^{\prime }\mathbf{p}}(t)=-\frac{g_{s}g_{v}}{2S}\int V\left( 
\mathbf{q}\right) \left( 1+e^{i2\zeta \left[ \vartheta (\mathbf{p}+\frac{e}{c%
}\mathbf{A}(t))-\vartheta (\mathbf{p}^{\prime }+\frac{e}{c}\mathbf{A}(t))%
\right] }\right)
\end{equation*}%
\begin{equation}
\times e^{-\frac{i}{\hbar }\int_{0}^{\tau }\left[ \left( \frac{1}{2m_{\ast }}%
\left( p_{x}^{\prime }+\frac{e}{c}A_{x}\right) ^{2}+\frac{p_{y}^{\prime 2}}{%
2m_{\ast }}-\mathcal{E}^{\prime }\right) -\left( \frac{1}{2m_{\ast }}\left(
p_{x}+\frac{e}{c}A_{x}\right) ^{2}+\frac{p_{y}^{2}}{2m_{\ast }}-\mathcal{E}%
\right) \right] dt}.  \label{7}
\end{equation}%
In accordance with Eq. (\ref{7}) the amplitude $T_{\mathbf{p}^{\prime }%
\mathbf{p}}(t)$ can be expressed in the following form:%
\begin{equation}
T_{\mathbf{p}^{\prime }\mathbf{p}}(t)=\frac{1}{S}e^{-\frac{i}{\hbar }\left( 
\mathcal{E}^{\prime }-\mathcal{E}\right) t}B(t),  \label{7b}
\end{equation}%
where the time-depended function%
\begin{equation*}
B(t)=-\frac{g_{s}g_{v}}{2}V\left( \mathbf{q}\right)
\end{equation*}%
\begin{equation*}
\times \left( 1+e^{i2\zeta \left[ \vartheta (\mathbf{p}+\frac{e}{c}\mathbf{A}%
(t))-\vartheta (\mathbf{p}^{\prime }+\frac{e}{c}\mathbf{A}(t))\right]
}\right)
\end{equation*}%
\begin{equation}
\times e^{-\frac{i}{\hbar }\int_{0}^{\tau }\left[ \left( \frac{1}{2m_{\ast }}%
\left( p_{x}^{\prime }+\frac{e}{c}A_{x}\right) ^{2}+\frac{p_{y}^{2}}{%
2m_{\ast }}-\mathcal{E}^{\prime }\right) -\left( \frac{1}{2m_{\ast }}\left(
p_{x}+\frac{e}{c}A_{x}\right) ^{2}+\frac{p_{y}^{2}}{2m_{\ast }}-\mathcal{E}%
\right) \right] dt}.  \label{17}
\end{equation}%
Making a Fourier transformation of the function $B(t)$\ (\ref{17}) over $t$%
,\ using the known relations%
\begin{equation}
B(t)=\sum\limits_{n=-\infty }^{\infty }e^{-in\omega t}\widetilde{B}_{\mathbf{%
p}^{\prime }\mathbf{p}}^{\left( n\right) },  \label{19}
\end{equation}%
\begin{equation}
\widetilde{B}_{\mathbf{p}^{\prime }\mathbf{p}}^{\left( n\right) }=\frac{%
\omega }{2\pi }\int_{0}^{2\pi /\omega }e^{in\omega t}B(t)dt,  \label{20}
\end{equation}%
we can write the SB amplitude $T_{\mathbf{p}^{\prime }\mathbf{p}}(t)$ (\ref%
{7}) as%
\begin{equation*}
T_{\mathbf{p}^{\prime }\mathbf{p}}(t)=\frac{1}{S}\sum\limits_{n=-\infty
}^{\infty }e^{-in\omega t}\widetilde{B}_{\mathbf{p}^{\prime }\mathbf{p}%
}^{\left( n\right) }e^{-\frac{i}{\hbar }\left( \mathcal{E}^{\prime }-%
\mathcal{E}\right) t}=
\end{equation*}%
\begin{equation}
\frac{1}{S}\sum\limits_{n=-\infty }^{\infty }e^{-in\omega t}\widetilde{M}_{%
\mathbf{p}^{\prime }\mathbf{p}}^{\left( n\right) }.  \label{15}
\end{equation}

To present the microscopic quantum theory of the multiphoton
inverse-bremsstrahlung absorption of coherent EM radiation in doped BG we
need to solve the Liouville-von Neumann equation for the density matrix $%
\widehat{\rho }$:%
\begin{equation}
\frac{\partial \widehat{\rho }}{\partial t}=\frac{i}{\hbar }\left[ \widehat{%
\rho },\mathcal{H}_{0}+\mathcal{H}_{sb}(t)\right]  \label{22a}
\end{equation}%
with the initial condition 
\begin{equation}
\widehat{\rho }(0)=\widehat{\rho }_{g}.  \label{23a}
\end{equation}%
Before the interaction with EM wave, it is assumed that the system of doped
BG quasiparticles was an ideal Fermi gas in equilibrium (thermal and
chemical) with a reservoir. Thus the density matrix $\widehat{\rho }_{g}$ of
the grand canonical ensemble is:%
\begin{equation}
\widehat{\rho }_{g}=\exp \left[ \frac{1}{T_{e}}\left( W+\int d\Phi _{\mathbf{%
p}}\left( \mu -\mathcal{E}\right) \widehat{a}_{\mathbf{p}}^{+}\widehat{a}_{%
\mathbf{p}}\right) \right] .  \label{24a}
\end{equation}%
In Eq. (\ref{24a}) $\mu $ is the chemical potential, $T_{e}$\ is the
electrons temperature in energy units, $W$ is the grand potential. The
initial single-particle density matrix in momentum space will be diagonal,
and we will have the fermionic distribution:%
\begin{equation*}
\rho \left( \mathbf{p}_{1},\mathbf{p}_{2},0\right) =Tr\left( \widehat{\rho }%
_{g}\widehat{a}_{\mathbf{p}_{2}}^{+}\widehat{a}_{\mathbf{p}_{1}}\right) =
\end{equation*}%
\begin{equation}
f(P_{1})\frac{(2\pi \hbar )^{2}}{S}\delta \left( \mathbf{p}_{1}-\mathbf{p}%
_{2}\right) ,  \label{25a}
\end{equation}%
where%
\begin{equation}
f(P_{1})=\frac{1}{\exp \left( \frac{\mathcal{E}_{1}-\mu }{T_{e}}\right) +1}.
\label{26a}
\end{equation}%
Here $Tr(\widehat{a}\widehat{b}\widehat{c})$ is the trace of a product of
the matrices functions $\widehat{a},\widehat{b},\widehat{c}$. Within the
Born approximation, we consider SB interaction Hamiltonian $\mathcal{H}%
_{sb}(t)$ as a perturbation. So, we expand the density matrix as%
\begin{equation}
\widehat{\rho }=\widehat{\rho }_{g}+\widehat{\rho }_{1},  \label{27a}
\end{equation}%
taking into account the relations%
\begin{equation*}
\left[ \widehat{a}_{\mathbf{p}^{\prime }}^{+}\widehat{a}_{\mathbf{p}},%
\widehat{\rho }_{g}\right] =\left( 1-e^{\frac{1}{T_{e}}\left( \mathcal{E}%
^{\prime }-\mathcal{E}\right) }\right) \widehat{\rho }_{g}\widehat{a}_{%
\mathbf{p}^{\prime }}^{+}\widehat{a}_{\mathbf{p}},
\end{equation*}%
\begin{equation*}
\left[ \widehat{\rho }_{g},\mathcal{H}_{0}\right] =0,
\end{equation*}%
for $\widehat{\rho }_{1}$ we obtain:%
\begin{equation*}
\widehat{\rho }_{1}=-\frac{i}{\hbar }\int\limits_{0}^{t}dt^{\prime }\int
d\Phi _{\mathbf{p}}\int d\Phi _{\mathbf{p}^{\prime }}B(t^{\prime })
\end{equation*}%
\begin{equation}
\times e^{\frac{i}{\hbar }\left( t^{\prime }-t\right) \left( \mathcal{E}%
^{\prime }-\mathcal{E}\right) }\left( 1-e^{\frac{1}{T_{e}}\left( \mathcal{E}%
^{\prime }-\mathcal{E}\right) }\right) \widehat{\rho }_{g}\widehat{a}_{%
\mathbf{p}^{\prime }}^{+}\widehat{a}_{\mathbf{p}}.  \label{28a}
\end{equation}%
The energy absorption rate of electrons due to inverse bremsstrahlung can be
presented as%
\begin{equation}
\frac{\partial E}{\partial t}=Tr\left( \widehat{\rho }_{1}\frac{\partial 
\mathcal{H}_{sb}(t)}{\partial t}\right) .  \label{29a}
\end{equation}%
It is convenient to represent the rate of inverse bremsstrahlung absorption
via the mean number of absorbed photons per impurity ion, per unit time:%
\begin{equation}
\frac{dN_{abs}}{dt}=\frac{1}{\hbar \omega N_{i}}\frac{\partial E}{\partial t}%
,  \label{30a}
\end{equation}%
where $N_{i}$ is the number of impurity ions in the interaction region.

Taking into account the decomposition relation:%
\begin{equation*}
\left( 1-e^{\frac{1}{T_{e}}\left( \mathcal{E}_{1}-\mathcal{E}_{2}\right)
}\right) Tr\left( \widehat{\rho }_{g}\widehat{a}_{\mathbf{p}_{1}}^{+}%
\widehat{a}_{\mathbf{p}_{2}}\widehat{a}_{\mathbf{p}_{3}}^{+}\widehat{a}_{%
\mathbf{p}_{4}}\right) =
\end{equation*}%
\begin{equation}
\left( 1-e^{\frac{1}{T_{e}}\left( \mathcal{E}_{1}-\mathcal{E}_{2}\right)
}\right) f_{1}\left( 1-f_{2}\right) ,  \label{31a}
\end{equation}%
and making the some calculations using the relations (\ref{28a})-(\ref{31a}%
), (\ref{11a}), (\ref{7b}), (\ref{17}) for large $t$\ we obtain:%
\begin{equation}
\frac{dN_{abs}}{dt}=\sum\limits_{n=1}^{\infty }\frac{dN_{abs}\left( n\right) 
}{dt},  \label{31aa}
\end{equation}%
where the partial absorption rates have the following forms:%
\begin{equation*}
\frac{dN_{abs}\left( n\right) }{dt}=\frac{4\pi g_{s}g_{\mathrm{v}}n}{\hbar
N_{i}S^{2}}\int \int d\Phi _{\mathbf{p}}d\Phi _{\mathbf{p}^{\prime
}}\left\vert V\left( \mathbf{q}\right) \right\vert ^{2}\left\vert \widetilde{%
M}_{\mathbf{p}^{\prime }\mathbf{p}}^{\left( n\right) }\right\vert ^{2}
\end{equation*}%
\begin{equation*}
\times \delta \left( \mathcal{E}^{\prime }-\mathcal{E}+n\hbar \omega \right)
\left( 1-e^{\frac{1}{T_{e}}\left( \mathcal{E}^{\prime }-\mathcal{E}\right)
}\right)
\end{equation*}%
\begin{equation}
\times f\left( \mathcal{E}^{\prime }\right) \left( 1-f\left( \mathcal{E}%
\right) \right) ,  \label{32a}
\end{equation}%
where 
\begin{equation*}
\left\vert \widetilde{M}_{\mathbf{p}^{\prime }\mathbf{p}}^{\left( n\right)
}\right\vert ^{2}=\left\vert \int_{0}^{T}d\left( \frac{t}{T}\right) \left(
1+e^{i2\zeta \left[ \vartheta (\mathbf{p}_{0}+\frac{e}{c}\mathbf{A}%
(t))-\vartheta (\mathbf{p}+\frac{e}{c}\mathbf{A}(t))\right] }\right) \exp
(in\omega t)\right.
\end{equation*}%
\begin{equation}
\times \left. e^{-\frac{i}{\hbar }\int_{0}^{\tau }\left[ \left( \frac{1}{%
2m_{\ast }}\left( p_{x}+\frac{e}{c}A_{x}\right) ^{2}+\frac{p_{y}^{2}}{%
2m_{\ast }}-\mathcal{E}\right) -\left( \frac{1}{2m_{\ast }}\left( p_{0x}+%
\frac{e}{c}A_{x}\right) ^{2}+\frac{p_{0y}^{2}}{2m_{\ast }}-\mathcal{E}%
_{0}\right) \right] dt^{\prime }}\right\vert ^{2}.  \label{33a}
\end{equation}%
The obtained expression for the absorption rate is applicable to arbitrary
polarization and intensity of the pump wave. It is true for a grand
canonical ensemble and positive only. The Dirac $\delta $ function expresses
the quadratic dispersion law in the SB process. This parabolic dispersion $%
\mathcal{E}^{\prime }(p)\mathcal{=}\sigma \left( p_{x}^{2}+p_{y}^{2}\right)
/2m_{\ast }+n\hbar \omega $ applies only for small values of $p$ satisfying $%
p/\hbar \ll \gamma _{1}/\left( \hbar \mathrm{v}_{F}\right) $ and for carrier
densities (or equivalently, the impurity densities) smaller than $5\times
10^{12}\mathrm{cm}^{-2}$. In the opposite limit, $p/\hbar \gg \gamma
_{1}/\left( \hbar \mathrm{v}_{F}\right) $ and for carrier densities larger
than the last, we get a linear band dispersion, $\varepsilon ^{\prime
}(p)=\sigma \hbar p\mathrm{v}_{F}+n\hbar \omega $, just as in a case of
intrinsic SG \cite{nanop2}. In the following with the help of (\ref{33a})
one can calculate the nonlinear inverse-bremsstrahlung absorption rate for
degenerate quantum plasma state - the BG chiral particles with the
distribution function given by the Eq. (\ref{26a}). 
\begin{figure}[tbp]
\centering{\includegraphics[width=.51\textwidth]{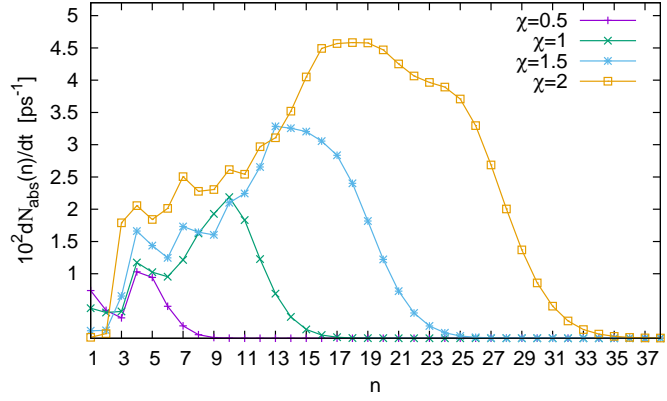}}
\caption{(Color online) Envelope of partial rate $dN_{abs}\left( n\right)
/dt $ of inverse bremsstrahlung absorption vs the mean number of absorbed
photons by per ion, per unit time (in \textrm{ps}$^{-1}$) for linear
polarization of EM wave in doped graphene is shown for various wave
intensities at $\protect\varepsilon \equiv \hbar \protect\omega =0.01$ 
\textrm{eV, }$T_{e}=0.1\protect\varepsilon _{F}$, and $\protect\varepsilon %
_{F}\simeq \protect\mu =20\protect\varepsilon $.}
\end{figure}
\begin{figure}[tbp]
\centering{\includegraphics[width=.51\textwidth]{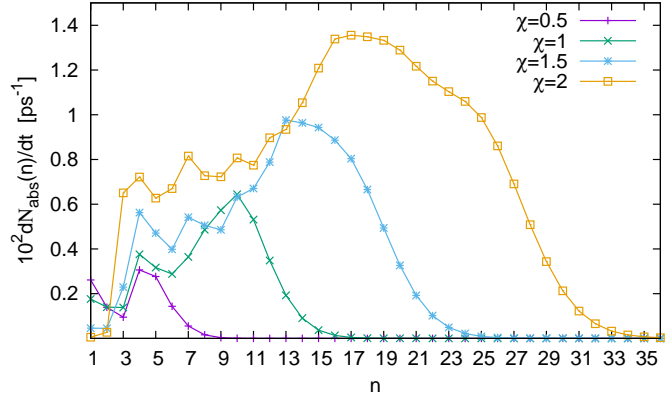}}
\caption{(Color online) Envelope of partial rate $dN_{abs}\left( n\right)
/dt $ of inverse bremsstrahlung absorption vs the mean number of absorbed
photons for photon energy $\protect\varepsilon =0.005$\ $\mathrm{eV.}$}
\end{figure}

\section{Numerical results for SB absorption coefficient for the screened
Coulomb potential of impurity ions in BG}

Now we utilize Eq. (\ref{33a}) in order to obtain the inverse-bremsstrahlung
absorption coefficient in particular case of SB process on a screened
Coulomb potential of impurity ions in BG. One we needs to concretize the
impurity ionic potential $V\left( \mathbf{q}\right) $, using an analytic
form for Coulomb screening in BG \cite{Screening}, \cite%
{Ando1982,Sarma2011,Kaikai Xu17}. So, the Fourier transform $V\left( \mathbf{%
q}\right) =\int \varphi (\mathbf{r})e^{-i\mathbf{qr}}d\mathbf{r}$ of a
charged impurity center potential\textbf{\ }can be written as:%
\begin{equation}
\left\vert V\left( \mathbf{q}\right) \right\vert ^{2}=N_{i}\frac{4\pi
^{2}e^{4}}{\kappa ^{2}q^{2}\epsilon ^{2}\left( q\right) }.  \label{29}
\end{equation}%
Here the screening term $\epsilon \left( q\right) $ ($q=\left\vert \mathbf{q}%
\right\vert $) is the 2D finite temperature static dielectric function in
random phase approximation (RPA) appropriate for BG, given by the formula 
\cite{Screening}, \cite{Sarma2011}:%
\begin{equation}
\epsilon \left( q\right) =1+\frac{q_{s}}{q}\left[ g\left( q\right) -f\left(
q\right) \Theta \left( q-2k_{F}\right) \right] .  \label{30}
\end{equation}%
Thee $k_{F}=\sqrt{2m_{\ast }\varepsilon _{F}}/\hbar $ is 2D Fermi wave
vector in BG case, $q_{s}=k_{TF}/k_{F}=4m_{\ast }e^{2}\log 4/\left( \kappa
\hbar ^{2}\right) $ ($\sim n^{-1/2}$ for BG \cite{Sarma2011}) is the 2D
Thomas-Fermi screening wave vector given by $k_{TF}$ \cite{Ando1982} scaled
on $k_{F}$; and $\kappa $ is the background lattice dielectric constant of
the system. The function $\Theta \left( q-2k_{F}\right) $ is the step
function. The functions $g\left( q\right) $, $f\left( q\right) $ are defined
by the formulas \cite{Sarma2011}:%
\begin{equation}
g\left( q\right) =\frac{1}{2}\sqrt{4+\overline{q}^{4}}-\log \left[ \frac{1+%
\sqrt{1+\overline{q}^{4}/4}}{2}\right] ,  \label{31}
\end{equation}%
\begin{equation}
f\left( q\right) =\left( 1+\frac{\overline{q}^{2}}{2}\right) \sqrt{1-\frac{4%
}{\overline{q}^{2}}}+\log \left[ \frac{\overline{q}-\sqrt{\overline{q}^{2}-4}%
}{\overline{q}+\sqrt{\overline{q}^{2}-4}}\right] ,  \label{31b}
\end{equation}%
where $\overline{q}=q/k_{F}$. This usual 2D dispersion or the static
screening \cite{Screening} one is the particular case of the wave vector
dependent plasmon dispersion and the wave frequency-dependent screening
function case \cite{Sarma2007}. For the last dielectric function and
screening show very different behavior than in intrinsic graphene case, and
the plasmons creation will be significant.
 
\begin{figure}[tbp]
\centering{\includegraphics[width=.51\textwidth]{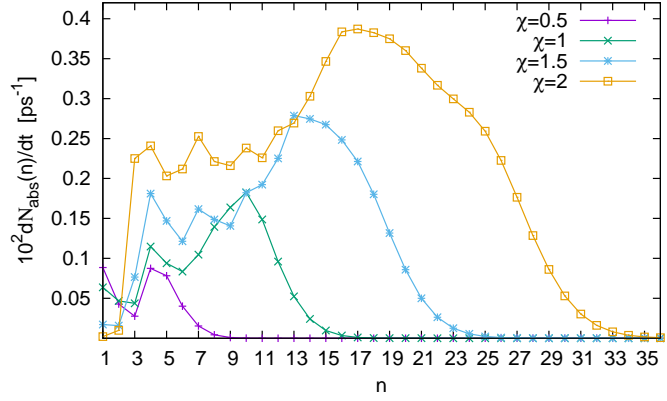}}
\caption{(Color online) Envelope of partial rate $dN_{abs}\left( n\right)
/dt $ of inverse bremsstrahlung absorption vs the mean number of absorbed
photons for photon energy $\protect\varepsilon =0.0025$\ $\mathrm{eV.}$}
\end{figure}

Taking into account Eqs. (\ref{33a}),(\ref{29}), and integrating in Eq. (\ref%
{32a}) over $\mathcal{E}^{\prime }$, we will obtain the following relation
for the partial absorption rates $dN_{abs}\left( n\right) /dt$ of SB process:%
\begin{equation*}
\frac{dN_{abs}\left( n\right) }{dt}=\frac{g_{s}g_{\mathrm{v}}n}{\pi \hbar }%
\frac{\left( \hbar q_{s}\right) ^{2}}{16\log ^{2}4}\int_{\mathcal{E}+n\hbar
\omega }d\mathcal{E}\int \int d\theta d\theta ^{\prime }\frac{\left\vert 
\widetilde{M}_{\mathbf{p}^{\prime }\mathbf{p}}^{\left( n\right) }\right\vert
^{2}}{\left( \hbar q\right) ^{2}\epsilon ^{2}\left( q\right) }
\end{equation*}%
\begin{equation}
\times \left( 1-e^{-\frac{n\hbar \omega }{T_{e}}}\right) f\left( \mathcal{E}%
-n\hbar \omega \right) \left( 1-f\left( \mathcal{E}\right) \right) .
\label{exact}
\end{equation}%

\begin{figure}[tbp]
\centering{\includegraphics[width=.51\textwidth]{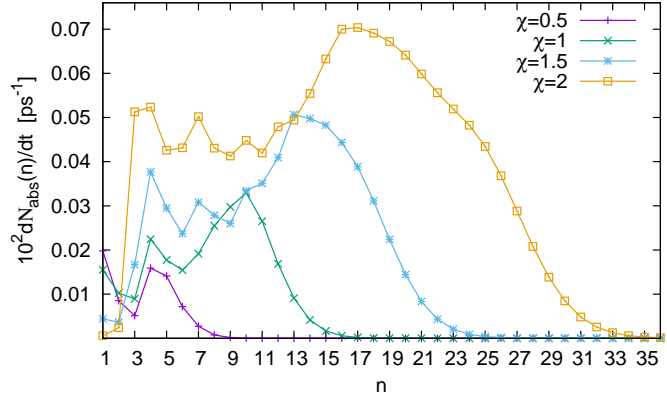}}
\caption{(Color online) Envelope of partial rate $dN_{abs}\left( n\right)
/dt $ of inverse bremsstrahlung absorption vs the mean number of absorbed
photons for photon energy $\protect\varepsilon =0.001$\ $\mathrm{eV.}$}
\end{figure}

For the consideration of numerical results it is convenient to represent the
differential cross-sections of SB on the charged impurities in the form of
dimensionless quantities. For the dimensionless rates $TdN_{abs}\left(
n\right) /dt$ in the field of linearly polarized EM\ wave with the
dimensionless vector potential $\overline{\mathbf{A}}(t)=-\widehat{\mathbf{x}%
}{\chi }\sin (2\pi \tau )$ we have $\mathrm{:}$%
\begin{equation*}
T\frac{dN_{abs}\left( n\right) }{dt}=g_{s}g_{v}n\frac{\left( \overline{\hbar
q_{s}}\right) ^{2}}{8\log ^{2}4}\int_{\overline{\mathcal{E}}+n}d\overline{%
\mathcal{E}}\int \int d\theta _{\mathbf{p}}d\theta _{\mathbf{p}^{\prime }}%
\frac{\left\vert \widetilde{M}_{\mathbf{p}^{\prime }\mathbf{p}}^{\left(
n\right) }\right\vert ^{2}}{\left\vert \hbar \overline{\mathbf{q}}%
\right\vert ^{2}\epsilon ^{2}\left( \left\vert \overline{\mathbf{q}}%
\right\vert \right) }
\end{equation*}%
\begin{equation}
\times \left( 1-e^{-\frac{n}{\overline{T_{e}}}}\right) f\left( \overline{%
\mathcal{E}}-n\right) \left( 1-f\left( \overline{\mathcal{E}}\right) \right)
,  \label{exact2}
\end{equation}%
where%
\begin{equation*}
\left\vert \overline{M}_{\mathbf{p}^{\prime }\mathbf{p}}^{\left( n\right)
}\right\vert ^{2}=\left\vert \int_{0}^{1}d\tau \left( 1+e^{i2\zeta \left[
\vartheta (\overline{\mathbf{p}}_{0}-{\chi }\sin (2\pi \tau ))-\vartheta (%
\overline{\mathbf{p}}-{\chi }\sin (2\pi \tau ))\right] }\right) \right.
\end{equation*}%
\begin{equation*}
\times \exp \left\{ i2\pi n\tau -\pi i\int_{0}^{\tau }\left[ \left( \left( 
\overline{p}_{x}^{\prime }-{\chi }\sin (2\pi \tau ^{\prime })\right) ^{2}+%
\overline{p}_{y}^{\prime 2}-2\overline{\mathcal{E}}^{\prime }\right) \right.
\right.
\end{equation*}%
\begin{equation}
\left. \left. \left. -\left( \left( \overline{p}_{x}-{\chi }\sin (2\pi \tau
^{\prime })\right) ^{2}+\overline{p}_{y}^{2}-2\overline{\mathcal{E}}\right) %
\right] d\tau ^{\prime }\right\} \right\vert ^{2}.  \label{333}
\end{equation}%
In Eq. (\ref{exact2}) the dimensionless momentum, energy, screening vector,
time are defined as follows:%
\begin{equation}
\overline{p}_{x,y}=\frac{p_{x,y}}{\sqrt{m_{\ast }\hbar \omega }},\overline{%
\mathcal{E}}=\frac{\mathcal{E}}{\hbar \omega },  \label{32}
\end{equation}

\begin{equation}
\overline{\hbar q}_{s}=\frac{\hbar q_{s}}{\sqrt{m_{\ast }\hbar \omega }}%
,d\tau =\frac{dt}{T}.  \label{34}
\end{equation}

For numerical analysis of SB cross-sections in BG the integration in Eq. (%
\ref{exact2}) over scattering angles $d\theta _{\mathbf{p}},d\theta _{%
\mathbf{p}^{\prime }}$, will be done by numerical calculations by the
fourth-order Runge-Kutta method. We assume Fermi energy $\varepsilon
_{F}\simeq \mu =20\hbar \omega $ ($\varepsilon _{F}\gg \hbar \omega $),
electrons temperature $T_{e}=0.1\varepsilon _{F}$, coherent EM linearly
polarized radiation field with frequency in terahertz or, close to it,
near-infrared  ($1.24-124$ $\mathrm{meV}$) for all calculations. As was
mentioned in Sec. II the band dispersion is quadratic and only the lowest
subband is occupied \cite{2e,3a,Screening}, and the RPA theory apply in the
density domain $10^{10}$ $\mathrm{cm}^{-2}<n<5\times 10^{12}$ $\mathrm{cm}%
^{-2}$. In particular case for a homogeneous carrier density $n=10^{11}$ $%
\mathrm{cm}^{-2}$ the dimensionless effective screening vector $q_{s}$
defined by the formula \cite{Sarma2011}: 
\begin{equation*}
q_{s}=\frac{54.8}{\sqrt{n\times 10^{-10}\mathrm{cm}^{2}}},
\end{equation*}%
The dielectric environment constant is taken to be $\kappa =2.5$ for an
impurity strength in the presence of the $\mathrm{SiO}_{2}$ substrate \cite%
{BLG}. To reveal the peculiarities which can be associated with the chiral
nature of BG and its parabolic dispersion, the comparison with the SG case
has been made.

The envelope of the partial rate of inverse-bremsstrahlung absorption in
doped BG is shown for different wave intensities in Fig. 1-4 for photon
energies $\varepsilon \equiv \hbar \omega =0.01$ \textrm{eV }($\lambda
=1.24\times 10^{-2}$ $\mathrm{cm}$), $\varepsilon =0.005$ \textrm{eV }($%
\lambda =2.48\times 10^{-2}$ $\mathrm{cm}$), $\varepsilon =0.0025$ \textrm{%
eV }($\lambda =0.5\times 10^{-2}$ $\mathrm{cm}$), and $\varepsilon =0.001$ 
\textrm{eV }($\lambda =0.124$ $\mathrm{cm}$), respectively. As was expected
the results substantially depend on coherent radiation frequency and
intensity: $I_{\chi }=\chi ^{2}\times 6\times 10^{10}$ $\mathrm{Wcm}%
^{-2}(\hbar \omega /\mathrm{eV})^{3}$. As is seen from these figures, the
multiphoton effects become essential already at moderate radiation
intensities (with intensity parameter $\chi =0.5$). With the increase of
coherent radiation intensity, the contribution of the multiphoton absorption
processes become larger than the one-photon SB. The maximal multiphoton
cut-off number is achieved at $\overline{\mathcal{E}}+n\sim 35$ at the
intensity parameter $\chi =2$ for the considering frequencies of the pump
wave. Cut-off increases at the increase of the intensity at fixed wave
frequency in accordance with the parabolic dispersion of the $AB$ stacked
BG. In comparison with the doped intrinsic SG \cite{nanop2}, where for the
same interaction parameters one has a single maximum case, we can note, that
the envelopes of the partial rate of inverse bremsstrahlung absorption have
many characteristic maxima (minima) in the doped BG. From the comparison of Figs. 1-4
we can conclude that the partial rate of the absorption process in doped BG
increases with the increase of the photon energy at the fixed parameter $%
\chi $ of the pump wave intensity, or with the increase of $\chi $ at the
fixed photon energy.

The numerical results allow to demonstrate the dependence of the inverse
bremsstrahlung absorption rate on the coherent radiation intensity. So, in
Fig. 5 the total SB\ rate (\ref{31aa}) via the mean number of absorbed
photons per impurity ion, per unit time in doped graphene versus the
parameter ${\chi }$ for various photon energies is presented. In Fig. 6 we
plot the scaled absorption rate ${\chi }^{-2}dN_{abs}/dt$ versus the
intensity parameter ${\chi }$. Figure 6 demonstrates, that almost for all
considered photon energies the scaled SB rate is constant, which means that
even for intense radiation when the multiphoton effects are essential, the
absorption rate is proportional to wave intensity. The latter means that
with the increase of the wave intensity we approach to a quasiclassical regime
of interaction:  $\mathcal{E}^{\prime }-\mathcal{E}>>\hbar \omega $ and $%
\mathcal{E}^{\prime }-\mathcal{E}\sim \chi ^{2}$. For terahertz and
near-infrared photons (wavelengths from $3$ $\mathrm{\mu m}$ to $3$ $\mathrm{%
mm}$), the multiphoton interaction mode in BG can be achieved at intensities 
$I_{\chi }\sim 10^{3}$\ $\mathrm{Wcm}^{-2}-10^{5}$\ $\mathrm{Wcm}^{-2}$,
which are available in present \cite{laser1}, \cite{laser}. Thus, for these
intensities of the strong coherent radiation field, one can manipulate with
the dressed charged carriers transport properties of doped BG by the
mechanism of multiphoton SB absorption. 
\begin{figure}[tbp]
\centering{\includegraphics[width=.51\textwidth]{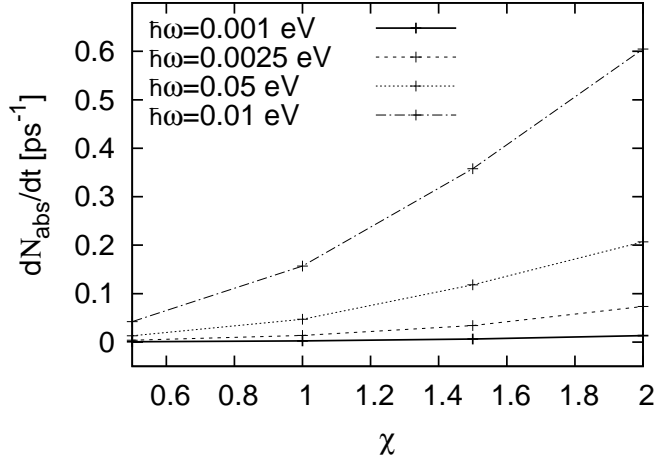}}
\caption{{}Total rate of inverse bremsstrahlung in doped graphene vs the
dimensionless parameter $\protect\chi $\ for the setup of Fig. 1 at various
photon energies $\protect\varepsilon $.}
\end{figure}
\begin{figure}[tbp]
\centering{\includegraphics[width=.51\textwidth]{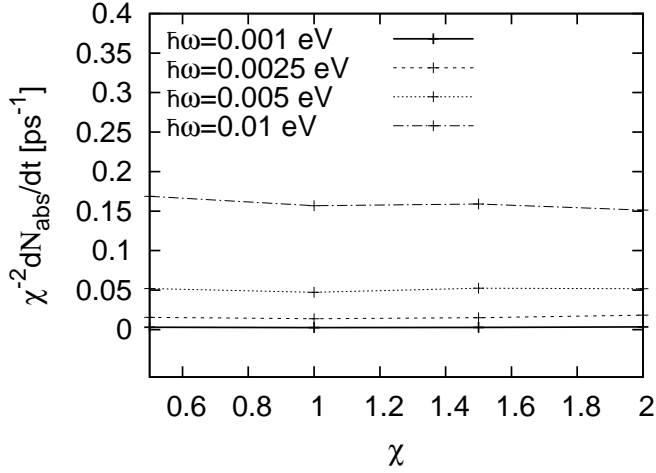}}
\caption{Total rates of the inverse bremsstrahlung absorption scaled to $%
\protect\chi ^{2}$\ vs the parameter $\protect\chi $.}
\end{figure}

\section{Conclusion}

The microscopic quantum theory of multiphoton SB absorption in doped BG ($AB$%
\ stacked) in the coherent strong EM radiation of arbitrary intensity has
been presented. The taken terahertz or near-infrared frequencies and Fermi
energies have allowed excluding the chiral particles' interband transitions.
We have solved the Liouville-von Neumann equation for the density matrix.
The external wave-field has been taken exactly. The charged impurity ions
arbitrary electrostatic potential is considered in the Born approximation.
These solutions for SB at the linear polarization of EM wave are used for
derivation of a relatively simple formula for the multiphoton SB absorption
rate. The chiral fermions in doped BG are represented by the grand canonical
ensemble. The obtained analytical formulas have been analyzed numerically
for screened Coulomb potential. The obtained results have an essentially
nonlinear dependence on the increase of the wave intensity. In comparison
with the SG case, the new behavior has been demonstrated, which may be
associated with its parabolic energy dispersion. In particular, there are
many characteristic maxima (minima) of the envelopes of the partial rate of inverse
bremsstrahlung absorption in a case of the doped BG. The multiphoton
absorption or emission processes play a significant role already at moderate
pump wave intensities $I_{\chi }\sim 10^{3}$\ $\mathrm{Wcm}^{-2}$. The
requirement of high intensity in the terahertz regime does not preclude the
use of standard terahertz lasers, which are available \cite{laser}.
Especially, of interest is the mid-infrared frequencies' range, where
Quantum Cascade lasers \cite{laser2} are readily available and can provide
higher powers. So, for such intensities, the multiphoton SB process opens
new channels for the wave absorption. It is shown that one can achieve the
efficient absorption coefficient by the SB process for pump wave frequencies in
considering domain.

\begin{acknowledgments}
The authors are deeply grateful to prof. H. K. Avetissian for permanent discussions and valuable recommendations.
This work was supported by the RA MES State Committee of Science.
\end{acknowledgments}

\end{document}